\documentclass[aps,prb,reprint,showpacs,amsmath,amssymb]{revtex4-1}

\usepackage{graphicx}
\usepackage{amsmath}
\usepackage{float}
\usepackage{float}

\begin{document}
\title{Ultrafast dynamics in the presence of antiferromagnetic correlations in electron-doped cuprate La$_{2-x}$Ce$_x$CuO$_{4\pm\delta}$}
\author{I. M. Vishik}
\altaffiliation[Current address: ]{University of California Davis, Department of Physics, Davis, CA, 95616, USA}
\affiliation {Massachusetts Institute of Technology, Department of Physics, Cambridge, MA, 02139, USA}
\author{F. Mahmood}
\altaffiliation[Current address: ]{Johns Hopkins University, Department of Physics and Astronomy, Baltimore, MD, 21218, USA}
\affiliation {Massachusetts Institute of Technology, Department of Physics, Cambridge, MA, 02139, USA}
\author{Z. Alpichshev}
\affiliation {Massachusetts Institute of Technology, Department of Physics, Cambridge, MA, 02139, USA}
\author{J. Higgins}
\affiliation{Center for Nanophysics and Advanced Materials, Department of Physics, University of Maryland, College Park, MD, 20472, USA}
\author{R. L. Greene}
\affiliation {Center for Nanophysics and Advanced Materials, Department of Physics, University of Maryland, College Park, MD, 20472, USA}
\author{N. Gedik}
\affiliation {Massachusetts Institute of Technology, Department of Physics, Cambridge, MA, 02139, USA}

\date{\today}

\begin{abstract}
We used femtosecond optical pump-probe spectroscopy to study the photoinduced change in reflectivity of thin films of the electron-doped cuprate La$_{2-x}$Ce$_x$CuO$_4$ (LCCO) with dopings of x$=$0.08 (underdoped) and x$=$0.11 (optimally doped).  Above T$_c$, we observe fluence-dependent relaxation rates which onset at a similar temperature that transport measurements first see signatures of antiferromagnetic correlations.  Upon suppressing superconductivity with a magnetic field, it is found that the fluence and temperature dependence of relaxation rates is consistent with bimolecular recombination of electrons and holes across a gap (2$\Delta_{AF}$) originating from antiferromagnetic correlations which comprise the pseudogap in electron-doped cuprates.  This can be used to learn about coupling between electrons and high-energy ($\omega>2\Delta_{AF}$) excitations in these compounds and set limits on the timescales on which antiferromagnetic correlations are static.
\end{abstract}

\maketitle

\section{Introduction}
Several families of unconventional superconductors share a canonical phase diagram whereby a superconducting dome emerges around the T$=$0 endpoint of an antiferromagnetic (AF) phase\cite{Scalapino:commonPairingUnconventionalSC}.  This may point to a common superconducting mechanism among organic, heavy fermion, pnictide, and cuprate superconductors \cite{Sachdev:AFHotspotPairing}, or it may highlight one factor which is generally favorable for the formation of superconductivity. In electron-doped cuprates, the AF phase is more robust than on the hole-doped side and may coexist with superconductivity to a greater degree \cite{Armitage:ElectronDopedReview}.  

In electron-doped cuprates which can be synthesized as bulk single crystals, neutron scattering is typically used to assess the onset of long-range AF order at the N\`{e}el temperature (T$_N$) and the finite correlation length at T$>$T$_N$.  Unlike hole-doped cuprates where the origin of the normal state pseudogap is still debated, the 'pseudogap' in electron-doped cuprates, first reported as a suppression of spectral weight in optics below a characteristic temperature T$^*$\cite{Onose:PG_NCCO} (sometimes called T$_W$\cite{Zimmers:SimulatingOptics}), is widely thought to originate from in-plane AF correlations with correlation length longer than the thermal deBroglie wavelength\cite{Motoyama:SpinCorrelationsNCCO,Fujita:LowEnergySpinFluctuationsPLCCO}.  The regime where AF correlations are present is marked by a single-particle gap of magnitude $\Delta_{AF}\approx 9 k_B T^*$ appearing at energy and momenta consistent with $(\pi, \pi)$ band folding\cite{Zimmers:SimulatingOptics,Matsui:FermiologyUDNCCO}.

La$_{2-x}$Ce$_x$CuO$_{4\pm\delta}$ (LCCO) has the highest maximum T$_c$ among electron-doped cuprates and exhibits superconductivity at lower doping, but it can only be stabilized as a thin film which prohibits characterization of antiferromagnetism via neutron scattering.  Instead, AF which is static on timescales of electron-relaxation times but not necessarily long-range has been identified via angular magnetoresistance (AMR)\cite{Jin:AFM_from_AMR_LCCO} with a characteristic onset temperature called T$_D$.  These transport measurements point to a phase diagram where AF coexists with superconductivity at optimal doping in LCCO (x=0.11).  On the other hand, low-energy $\mu$SR experiments indicated a more truncated AF phase, in which static order has a much more limited overlap with superconductivity, disappearing at x$\approx$0.08 \cite{Saadaoui:LCCOmuSR}.  These two experiments are not necessarily in conflict, as the two are sensitive to very different timescales.

Ultrafast optics can access the timescales intermediate between electron relaxation times probed by transport ($\approx$ 10-100 fs) and the typical Larmor frequency associated with the internal fields (0.1 $\mu$s).  Optical pump-probe experiments on optimally-doped Nd$_{2-x}$Ce$_x$CuO$_{4\pm\delta}$ (NCCO) have indicated that short-range AF correlations manifest in scaling behavior of the transient reflectivity, and that signatures of these correlations persist below T$_c$ and compete with superconductivity \cite{Hinton:NCCOUltrafast}.

Here we show time-resolved optical signatures of AF correlations in electron-doped cuprates, and use relaxation rates measured in these time-domain experiments to set limits on the AF correlation time and the strength of coupling between electrons and high energy ($\omega\geq 2\Delta_{AF}$) bosons. We have measured photoinduced reflectivity ($\Delta R/R$) in LCCO and extracted the temperature and pump-fluence dependence of the initial rate at which $\Delta R/R$ decays back to equilibrium. Above T$_c$ in both underdoped and optimally doped samples, AF correlations manifest as pump-fluence-dependent decay rates which are characteristic of a fully-formed gap in the density of states (DOS).  When superconductivity it suppressed with a magnetic field, it is revealed that the temperature and fluence dependence of initial decay rates is consistent with pairwise recombination of electron-hole excitations across $\Delta_{AF}$.

\section{Methods}
Ultrafast pump-probe experiments employ short pulses of light ($< 1$ ps) to create new electronic states, destroy electronic states, or make targeted excitations in solids.  The latter option is the focus of this study. These experiments consist of two pulses separated in time: the pump pulse perturbs the sample and the probe pulse studies the changes in electronic properties.  By varying the time delay between the pump and the probe, one can study how the non-equilibrium electronic state decays back to equilibrium, which can give information about the relaxation processes which are relevant to the important emergent phases in condensed matter systems.  

Experiments were performed with a Ti:sapphire oscillator lasing at  800 nm ($\hbar$$\omega$$=$1.55 eV) producing pulses 60 fs in duration. The repetition rate of the laser was reduced to 1.6 MHz with a pulse picker to mitigate against steady-state heating of the sample. Experiments were performed in two different cryostats depending if a magnetic field was applied or not.  For data in Figs. \ref{Fig 1} and \ref{Fig 2}, the sample was excited by a pump pulse of 70 $\mu$m FWHM diameter, and the fluence ($\Phi$) of the pump pulse was varied between 8.7 and 0.1 \begin{math}\mu J/\text{cm}^2\end{math}.  For data in Figs. \ref{Fig 3}, \ref{Fig 4}, and  \ref{Fig 5} the sample was excited by a pump pulse of 150 $\mu$m FWHM diameter, and $\Phi_{pump}$ was varied between 2.3 and 0.08 \begin{math}\mu J/\text{cm}^2\end{math}.  Results from the two experimental configurations are consistent.  This study accessed pump-fluences lower than some earlier studies \cite{Cao:QPRelaxationLCCO,Long:LCCOUltrafast} with the goal of making a small number of electronic excitations and not destroying the underlying order. The sample response was assessed through measurement of the normalized change in the reflectivity, $\Delta$R(t)$/$R, of a separate probe pulse which was focused on the same spot on the sample as the pump.  For data in Figs. \ref{Fig 1} and \ref{Fig 2}, $\Phi_{probe}=$0.9 $\mu$J$/$cm$^2$.  For data in Figs. \ref{Fig 3}, \ref{Fig 4}, and  \ref{Fig 5}, $\Phi_{probe}=$0.4 $\mu$J$/$cm$^2$.   The probe fluence was chosen to maximize the signal-to-noise ratio while avoiding steady-state-heating.  The c-axis-oriented LCCO was deposited directly on insulating (100) ​SrTiO$_3$ substrates by a pulsed laser deposition technique utilizing a KrF excimer laser. Two films with Ce concentrations of x$=$0.08 (underdoped) and 0.11 (optimally doped) were studied with T$_c$'s of 22K and 25K, respectively. The annealing process was optimized for each doping.

\section{Interpreting optical pump-probe data in electron-doped cuprates}

\subsection{What the pump does and what the probe measures}
In the experiments shown here, both the excitation (pump) and the reflectivity measurements (probe) are done at the same frequency, 1.55 eV (800 nm).  At the intensities and frequency used in these experiments, the pump initially makes excitations from occupied to unoccupied states.  In general, this initial excitation has higher energy than the lowest-lying excitations in the system.  Depending on the specific system under investigation, the lowest lying excitation might be the band gap of a semiconductor, the single particle gap due to a charge density wave (CDW) or a spin density wave (SDW), or the superconducting gap energy ($2\Delta$).  After the initial excitation, there is typically a cascade of relaxation processes driven by electron-electron or electron-phonon interactions which result in excitations primarily at the gap edge on sub-picosecond to picosecond timescales (Fig. \ref{Fig 7}(b)\cite{Lobo:PhotoinducedTimeResolvedElectrodynamicsBCS_SCs,Othonos:UltrafastDynamicsSemiconductorReview}.
A second assumption in the data interpretation is that the magnitude of the change in reflectivity at 800nm, $|\Delta R/R|$ is proportional to the number of gap-energy excitations, and the time evolution is hence related to the creation and annihilation of these excitations and/or their diffusion out of the excitation volume. Specifically, in electron-doped cuprates the excitations we will consider are photoexcited quasiparticles (broken Cooper pairs) and electron-hole excitations across $2\Delta_{AF}$.

It can be counterintuitive that excitations at energies orders of magnitude lower than the probe frequency of 1.55 eV can cause changes in reflectivity at the probe frequency.  This can be resolved by noting that frequency-dependent reflectivity is related to the complex dielectric function which in turn is related to both the real ($\sigma_1$) and imaginary ($\sigma_2$) part of the optical conductivity.  Because of the Kramers-Kronig relation between $\sigma_1(\omega)$ and $\sigma_2(\omega)$, which is an integral over all frequencies, changes to $\sigma_1$ at small frequency can ultimately manifest as changes, albeit small ones, to the reflectivity at much higher frequencies.  This has been modeled using parameters appropriate to hole doped cuprates \cite{Segre:thesis}, and it was shown that converting 1$\%$ of the condensate into quasiparticles at the gap edge (50 meV) can produce changes in reflectivity at 1.55 eV of order $10^{-4}$.

\subsection{Rothwarf-Taylor Model}
The relaxation dynamics in the presence of a small energy gap, such as the one due to superconductivity (SC), can be analyzed using the Rothwarf-Taylor (RT) model \cite{RT_model}.  These coupled differential equations consider the time evolution of populations of excited quasiparticles ($n \propto |\Delta R/R|$) which can recombine into Cooper pairs, emitting a boson, and the population of bosons (\textit{N}) which can break Cooper pairs to create quasiparticles.  Originally, these equations described dynamics in Bardeen-Cooper-Schrieffer (BCS) superconductors, in which pair formation/breaking is known to be facilitated by high-frequency ($\omega>2\Delta$) phonons.  However, this model has proved to be successful in other systems which have small energy gaps close to the Fermi level\cite{Kabanov:KineticsSC}, such as heavy fermions\cite{Demsar:HF_PP_review}, CDW systems\cite{Yusupov:PumpProbeTriTellurides2008}, and SDW systems\cite{Chia:QP_Relaxation_RT_SDW_2006}.  Additionally, the model permits pair breaking by bosons other than phonons.  The Rothwarf-Taylor equations are given by:
\begin{equation}
\frac{dn}{dt}=I_{qp}+2\gamma_{pc}N-\beta n^2
\end{equation}
\begin{equation}
\frac{dN}{dt}=I_{ph}+\frac{1}{2}\beta n^2-\gamma_{pc}N-(N-N_{eq})\gamma_{esc}
\end{equation}
where $I_{qp}$ is an external source of quasiparticles or low energy excitations in non-SC systems, $\gamma_{pc}$ is a pair creation rate of photoexcited quasiparticles or other low-energy excitations, $\beta$ is a bimolecular recombination constant, $I_{ph}$ is a source of non-equilibrium phonons or other bosons, $N_{eq}$ is the equilibrium population of pair-breaking phonons or other bosons, and $\gamma_{esc}$ is the rate at which these phonons/bosons either escape from the excitation volume or decay into lower frequency bosonic excitations.

Considering the case of a superconductor with phonon pair breaking, we discuss various regimes of the RT model.  In the unphysical case where $\gamma_{esc}=0$ and the rate of pair creation is comparable to the rate of recombination ($\gamma_{pc}\approx \beta n$),  the non-equilibrium population of photoexcited quasiparticles is maintained indefinitely because of detailed balance between populations of quasiparticles and phonons.  This is the so-called phonon bottleneck.  It should be noted that quasiparticle diffusion is not included in this formulation of the RT model.  In a regime where $\gamma_{esc}$ is finite and it still holds that $\gamma_{pc}\approx \beta n$, the phonon-bottleneck will still exist, although the $n$ will decrease over time, at a rate determined by the relative values\cite{Lobo:PhotoinducedTimeResolvedElectrodynamicsBCS_SCs} of $\gamma_{esc}$, $\gamma_{pc}$, and $\beta n$.  Finally, if $\gamma_{esc}\gg\gamma_{pc}$ or $\beta n\gg\gamma_{pc}$, the RT equations decouple and the time evolution of the quasiparticle population is determined by bimolecular recombination--that is, two quasiparticles (or an electron and a hole in the case of other small-gap excitations) recombining with one another: $\frac{dn}{dt}=-\beta n^2$.

One characteristic of bimolecular-recombination-dominated dynamics is that the decay rate, $\gamma_0$, depends on the number of gap-energy excitations, which is usually proportional to pump fluence.  Putting these these attributes together yields $\gamma_0\propto n$.  This characteristic linear relationship between excitation density and recombination rate has been observed in the SC state of hole-doped cuprates\cite{Segre:YBCOUltrafast,Gedik:YBCOPRB} and iron-based superconductors\cite{Torchinsky:PnictideUltrafast}.  When $\gamma_0$ is plotted as a function of pump fluence or a proportional quantity, the slope is related to the bimolecular recombination constant, $\beta$, which has physical origins.  For BCS superconductors in the dirty limit, $\beta$ is related to the ratio of the electron-phonon coupling function weighted by the phonon DOS at the gap energy ($\alpha^2(2\Delta)F(2\Delta)$) to the electronic density of states at the Fermi level ($N(0)$)\cite{Kaplan:QP_Phonon_lifetimes_SCs}.  An analogous physical origin can be derived in non-SC systems where non-equilibrium dynamics are dominated by bimolecular recombination, but in either case, another characteristic of such a system is that the slope of $\gamma_0$ vs $n$ is independent of temperature (T) for $k_B T$ well below $2\Delta$.

\subsection{Bimolecular recombination and data fitting}

In the regime where relaxation is set by bimolecular recombination, the excitation density as a function of time  is given by
\begin{equation}
\frac{\Delta R}{R}(t) \propto n(t)=\frac{n(0)}{1+\beta n(0)t}
\end{equation}
where $n(0)$ is the excitation density at the beginning of the recombination process, usually taken at the time when $|\Delta R/R|$ is maximum.
Additionally, one can account for the exponentially decreasing excitation density as a function of depth into the sample to yield a corrected transient reflectivity\cite{Gedik:YBCOPRB}:

\begin{equation}
\label{eqn fit bimolecular}
\Delta R=\frac{2\Delta R(0)}{\gamma_0 t}[1-\frac{\ln(1+\gamma_0 t)}{\gamma_0 t}]
\end{equation}

where $\Delta R(0)$ is the reflectivity change at the sample surface, $\gamma_0\equiv \beta n(0,0)$, and $n(0,0)$ is the excitation density at the sample surface.  $\Delta R(0)$ and $\gamma_0$ are free parameters in the fitting.

\subsection{Considerations specific to electron-doped cuprates}
\begin{figure}[!]
\includegraphics [type=eps,ext=.eps,read=.eps,clip, width=3.2 in]{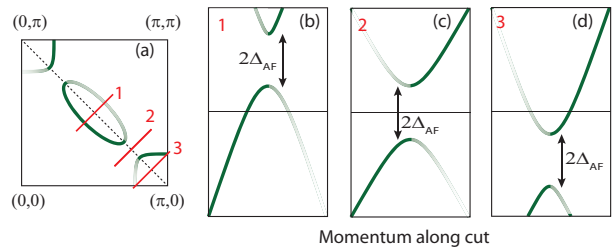}
\centering
\caption[Fig 1a] {\label{Fig 1a}  Schematic of fermiology and band structure in electron-doped cuprates in the presence of $\mathbf{q}=(\pi,\pi)$ reconstruction, widely attributed to AF order.  The SC gap is not shown.  (a) Fermi surface.  Dashed line marks AF zone boundary. (b)-(d) Schematic of dispersion along 3 cuts shown in (a).  Cut 2 is through the hot spot.}
\end{figure}
Dynamical mean field theory (DMFT) calculations have indicated that around 10$\%$ electron doping, the primary optical excitations permissible with a 1.5 eV pump are from the quasiparticle band into the upper Hubbard band\cite{Weber:CorrelationsBandStructureEdopedHdopedcuprate}. Additional but less probable possibilities exist for excitation from the low energy tail of the Zhang-Rice singlet band (centered at -2.1 eV in Ref. \onlinecite{Weber:CorrelationsBandStructureEdopedHdopedcuprate}) into the quasiparticle band.  The quasiparticle band is implicated in both superconductivity and AF, and at this doping, its momentum-integrated electronic DOS is split into two peaks separated by $2\Delta_{AF}$ on either side of the Fermi level ($E_F$).

Without momentum-integration, some momenta have metallic bands crossing $E_F$ on which a SC gap can open.
A schematic of the Fermi surface and low-energy band structure is shown in Fig. \ref{Fig 1a}.  In the regime where there is long-range AF order, the large hole-like fermi surface (FS) which is found in the overdoped regime, undergoes band folding at $\mathbf{q}=(\pi,\pi)$, yielding hole and electron pockets.  A gap ($\Delta_{AF}$) (sometimes called $\Delta_{PG}$ or $\Delta_{SDW}$) is opened everywhere that the original band crosses the folded band.  This gap is centered at $E_F$ only at the so-called hot-spot momentum, indicated in Fig. \ref{Fig 1a}(c); closer to the Brillouin zone center, $\Delta_{AF}$ appears above $E_F$ and closer to the Brillouin zone boundary, it appears below $E_F$\cite{Matsui:FermiologyUDNCCO}. As the AF correlation length becomes finite, the DOS inside the gap begins to fill in, and the gap is completely filled by $T=T^*$\cite{Park:SpinCorrelationsCalculation,Armitage:DopingDepNCCOARPES}.  The depression of DOS in the hotspots and the pseudogap observed by optics are commonly attributed to short-range AF, which is why we use this terminology, but it should be noted that other types of order, such as \textit{d}-density wave can produce the same effect\cite{Chakravarty:dDensityWaveCuprates}.

\section{Relaxation dynamics at B=0}
In this section, we present fluence- and temperature-dependent dynamics above T$_c$, as well as temperature dependence of normalized $\Delta R/R$ from low temperature to just above T$_c$.
\begin{figure*}[!]
\includegraphics [type=eps,ext=.eps,read=.eps,clip, width=6.0 in]{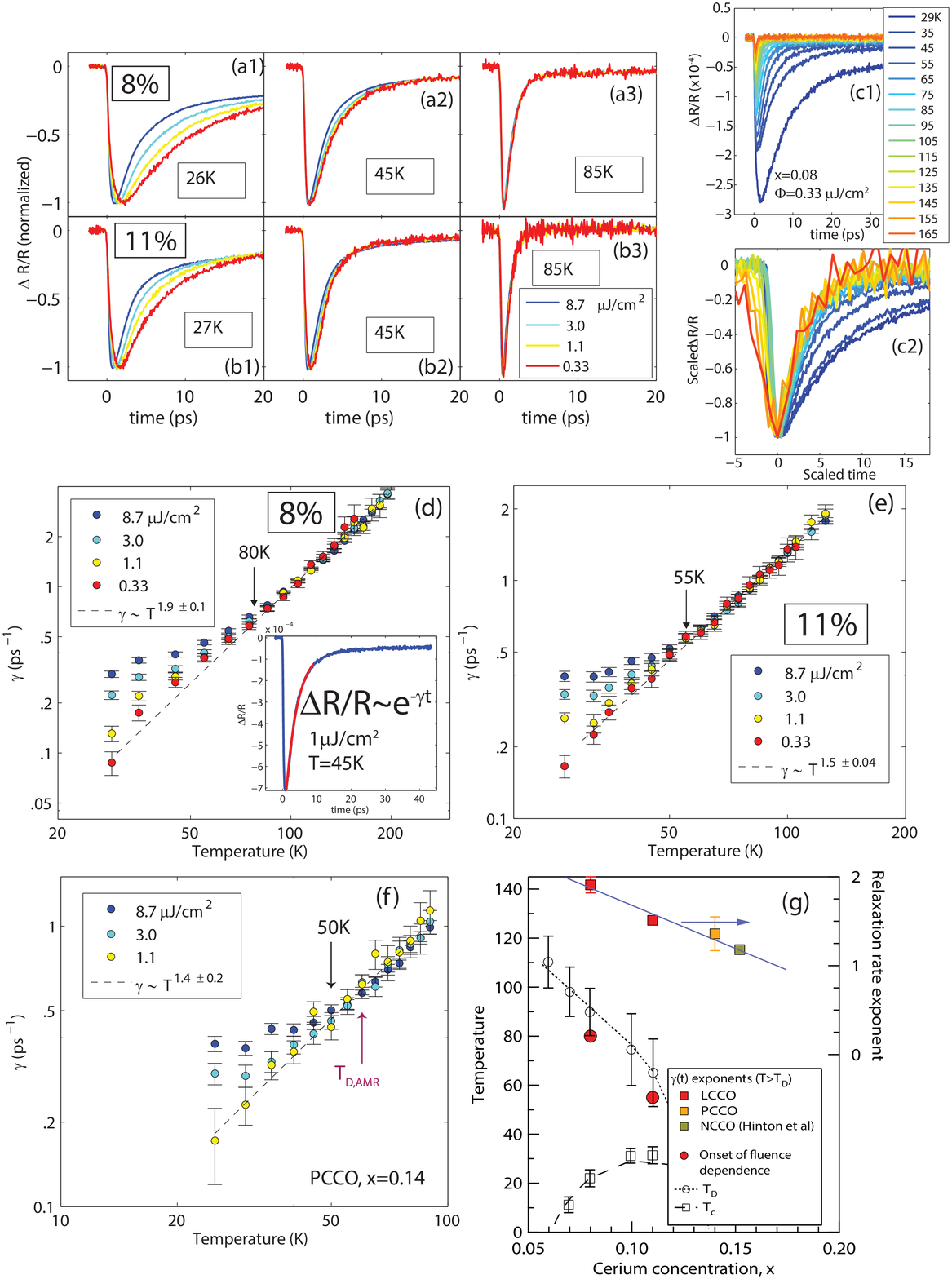}
\centering
\caption[Fig 1] {\label{Fig 1}  Temperature and pump-fluence dependence T$>$T$_c$. (a1)-(a3) x$=$0.08: normalized $|\Delta R/R|$ at different pump fluences and three representative temperatures above T$_c$. Non-overlap of curves at lower temperatures indicates fluence-dependent initial relaxation rates. (b1)-(b3) same for x$=$0.11. (c1) Temperature dependence of $\Delta R/R$ for x$=$0.08 at fixed fluence, $\Phi=0.33\mu J/cm^2$. (c2) Same data with x and y-axis scaled by the same factor. (d)-(e) Relaxation rate, $\gamma$ as a function of temperature and fluence for both dopings of LCCO above T$_c$.  Vertical arrow marks onset of systematic fluence dependence.  Dashed line shows power law fit in the temperature regime where there is no fluence dependence of $\gamma$.  Inset of (d) indicates that $\gamma$ is extracted from fit to a single exponential. (f) same for PCCO, x$=$0.14 (T$_c$$\approx$22K).  AMR value from Ref. \onlinecite{Yu:TransportAndAFMPCCO}.  For comparison, optics yields an estimated T* of 120K for this doping\cite{Zimmers:SimulatingOptics}. (g) Temperature scales in relation to transport experiments by Jin \textit{et al}\cite{Jin:LCCOTransportNature}.  Filled circles mark the temperatures derived in (d)-(f), and squares are relaxation rate exponents for LCCO, PCCO, and NCCO (Ref. \onlinecite{Hinton:NCCOUltrafast}) }
\end{figure*}

Prior time-resolved reflectivity studies on electron-doped cuprates observed a negative $\Delta R/R$ at 800 nm above T$_c$, and attributed this feature to either a collective mode of the pseudogap or in-plane AF correlations\cite{Hinton:NCCOUltrafast,Wang:2016UltrafastLCCO}.  We observe a similar feature, though our data favor the latter explanation.  We also expand on previous studies by analyzing its temperature and pump-fluence dependence in Fig. \ref{Fig 1}.   For this set of data, initial relaxation rates were derived from fitting to a single decaying exponent after the time when $\Delta R/R$ is maximum.  This fitting is chosen for this portion of the study to accommodate a wide temperature range of data which may encompass varying relaxation mechanisms.

The key finding is that $\Delta R/R$ in LCCO has fluence dependence above T$_c$, which is different from hole-doped cuprates where pump-fluence-dependence is absent above T$_c$\cite{Gedik:YBCOPRB}.  This fluence dependence is evident both from examining normalized data (Fig. \ref{Fig 1} (a)-(b)) and by fitting the data (Fig. \ref{Fig 1}(d)-(e)) to extract an initial relaxation rate.  The fluence dependence is absent at temperatures higher than where transport measurements demarcate AF correlations at T$_D$.  Similar behavior is observed in thin films of Pr$_{2-x}$Ce$_x$CuO$_{4\pm\delta}$ (PCCO)(Fig. \ref{Fig 1}(f)) indicating that these fluence-dependent relaxation dynamics are generic to electron-doped cuprates in the regime where AF correlations are observed by transport.

Above T$_D$, $\gamma$ increases with temperature following a power law (T$^\alpha$) with an exponent $\alpha$ between 1 and 2, whose value decreases with doping when several families are compared (Fig. \ref{Fig 1}(g)).  In metallic systems, the high-temperature relaxation rate has been connected to the strength of electron-phonon coupling\cite{Allen:TheoryThermalRelaxationElectronsMetals,Brorson:ElectronPhononCouplingRTSuperconductors_PP}, which in cuprates, has been shown to weaken with increasing doping\cite{Meevasana:CalcOD_caxis_charge_dynamics}.  A further possible relaxation mechanism in electron-doped cuprates is spin excitations, and the influence of magnetism progressively gets weaker as doping increases.  Finally, we note that our data lack the scaling behavior (Fig. \ref{Fig 1}(c)) reported in Ref. \onlinecite{Hinton:NCCOUltrafast}, likely because the dopings we consider are further from the quantum critical regime.

\begin{figure}[!]
\includegraphics [type=eps,ext=.eps,read=.eps,clip, width=3 in]{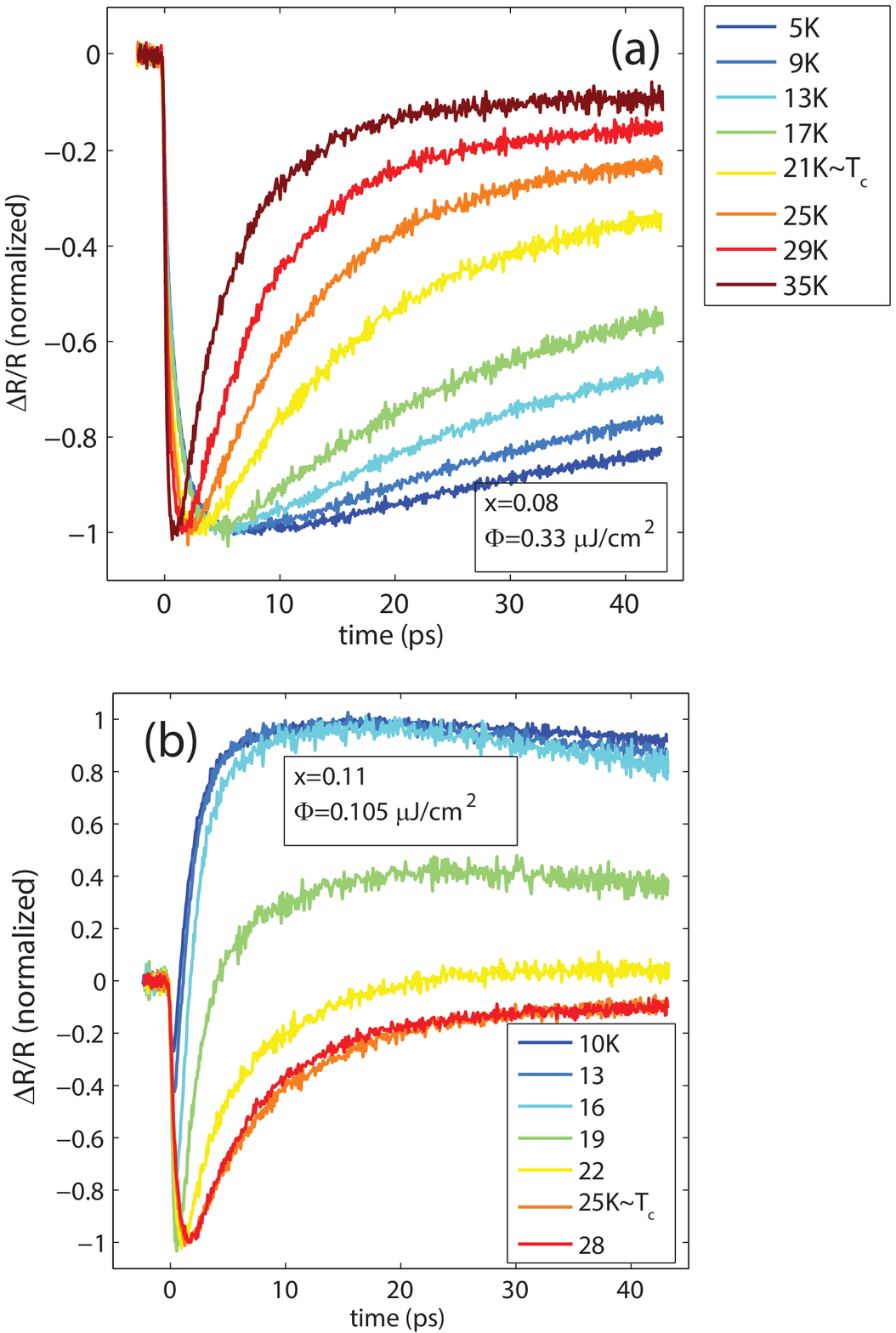}
\centering
\caption[Fig 2] {\label{Fig 2}  Temperature dependence across T$_c$. (a) x$=$0.08 (b) x$=$0.11. Both data sets have been normalized by dividing by the maximum of $|\Delta R/R|$ }
\end{figure}

Fig. \ref{Fig 2} shows temperature dependence across T$_c$ of normalized $\Delta R/R$ taken with pump fluence $<0.3\mu J/cm^2$.  The negative component in $\Delta R/R$ persists below T$_c$ in both dopings of LCCO.    In x$=$0.11, an additional component with positive sign in $\Delta R/R$ emerges below T$_c$, as was earlier observed in optimally doped NCCO\cite{Hinton:NCCOUltrafast}.  This positive component can be attributed to SC because it is absent above T$_c$.  In underdoped x$=$0.08 LCCO, $\Delta R/R$ remains negative below T$_c$.  Magnetic field in the next section clarifies that $\Delta R/R$ in x$=$0.08 consists of a negative AF component superimposed on a negative SC component.

\section{Isolating AF component with magnetic field}
In this section, we separate a SC component from an AF component in $\Delta R/R$ by applying a magnetic field close to H$_c2$.  From the fluence dependence of both, we can infer the primary mechanisms of relaxation for low energy excitations related to each state.

PCCO films with similar T$_c$ (23K) were used as guidance about H$_{c2}$ of our LCCO films at the measurement temperatures of 11K and 12K\cite{Fournier:Hc2PCCO}.  In particular, the field where in-plane resistivity, $\rho_{xx}$, begins to deviate from high-field linear magnetoresistance, marked as $H_{100}$ in Ref. \onlinecite{Fournier:Hc2PCCO}, has been shown to be the field where the condensate is extinguished.  For x$=$0.15 PCCO (T$_c$$=$23K), $H_{100}=7.5 T (7 T)$ at 11K (12K).

\begin{figure*}[!]
\includegraphics [type=eps,ext=.eps,read=.eps,clip, width=6 in]{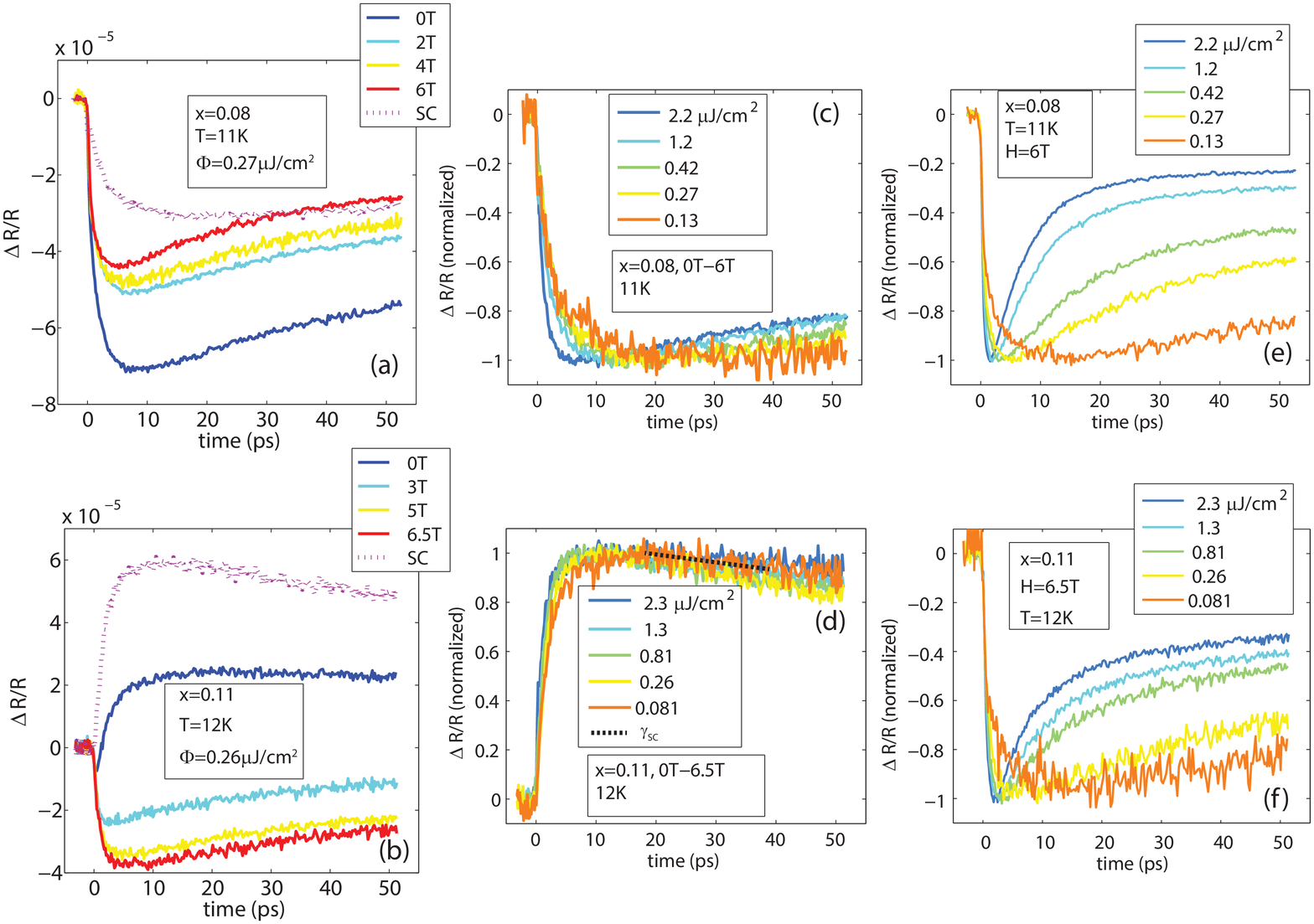}
\centering
\caption[Fig 3] {\label{Fig 3} Isolating two components with magnetic field.  (a) $\Delta R/R$ for x=0.08, taken at several magnetic field (B$||$c), 11K, and fixed pump fluence. Dashed curve shows 6T data subtracted from 0T data in order to isolate SC component. (b) Same for x$=$0.11, except SC component derived by subtracting 6.5T data from 0T data. (c)-(d) Fluence dependence of SC component for both dopings. Black dotted line in (d) indicates linear fit which is used to extract a relaxation rate for the SC component, $\gamma_{SC}$.(e)-(f) Fluence dependence of AF component, defined as $\Delta R / R$ at the maximum field studied, for both dopings.}
\end{figure*}

Fig. \ref{Fig 3} shows the effect of a magnetic fields up to 6.5T applied parallel to the \textit{c}-axis of the LCCO films.   As magnetic field increases, $\Delta R/R$ in x$=$0.11 becomes negative at all pump-probe delay times (Fig. \ref{Fig 3}(b)).  This confirms that the positive component is associated with SC in x$=$0.11, as suggested from temperature dependence.  In x$=$0.08, the magnitude of $\Delta R/R$ \textit{decreases} with increasing field (Fig. \ref{Fig 3}(a)), indicating that the SC component of $\Delta R/R$ has a negative sign at this doping and probe frequency.  In both cases, excitations across the SC gap and across the AF gap give distinct contributions to $\Delta R/R$.  Long-lived excitations across the AF gap may be the reason that near-infrared pumping was shown to extinguish the superconducting condensate at a much higher intensity than expected\cite{Beck:MeltingSC_PCCO_THz_NIR}.

A nominal superconducting component is extracted by subtracting the maximum-field data from the zero-field data, and this is shown in Fig. \ref{Fig 3}(c)-(d) and as the purple dashed lines in Fig. \ref{Fig 3}(a)-(b). This subtraction procedure is least fraught if there is no microscopic coexistence between superconductivity and the AF normal state, such that distinct regions of the sample contribute to AF and SC.  In the case of microscopic coexistence between AF and SC, two complications can arise: competition between the two orders, which would affect the magnitude of the gaps themselves, and \textit{populations} of photoexcited quasiparticles and excitations across $\Delta_{AF}$ being linked to one another.  In the former scenario, suppression of superconductivity can lead to enhancement of $\Delta_{AF}$, if the two orders are antagonistic, as is seen in iron-based superconductors\cite{Yi:CompetitionSDW_SC_FESC}.  However, for the case of electron-doped cuprates, the energy scales are quite different for SC and AF, with the former having a maximum magnitude of $\approx 3-5$ meV \cite{Matsui:nonMonatonicGapPLCCO,Niestemski:PLCCOTunnelingGapBoson,Diamant:DopingDepSCPCCO_PRB,Homes:OpticalDeterminationSCGap_PCCO} and the latter having a magnitude $> 80$ meV\cite{Armitage:ElectronDopedReview,Zimmers:SimulatingOptics}.  Thus, the effect of suppressing SC on the magnitude of $\Delta_{AF}$ is not expected to be as large as it is in systems where the competing orders have energy scales comparable to one another.  The second scenario where simple subtraction might not accurately yield the superconducting component is potentially more serious, and is applicable to excitations made close to the node of the superconducting order parameter, where $\Delta_{AF}$ opens \textit{above} the Fermi level (Fig. \ref{Fig 1a}(b)).  In this portion of the Brillouin zone, electron-hole recombination across $\Delta_{AF}$ may provide an additional source of photoexcited quasiparticles, making the extended Rothwarf-Taylor model a more appropriate starting point\cite{Springer:ExtendedRTModel}.  For that reason, we will keep the discussion of the superconducting component derived in the manner described above qualitative, and leave more sophisticated methods of component separation\cite{Alpichshev:SodiumIridate_TG} for a later study.

The first observation about the SC component of $\Delta R/R$ is that it changes sign between underdoped (negative) and optimal doping (positive).  It should be noted that optimally doped NCCO\cite{Hinton:NCCOUltrafast} and x$=$0.14 PCCO also have a SC component with a positive sign, so this observation may be generic to all families of electron-doped cuprates.  Previously, a sign change of $\Delta R/ R$ below T$_c$ in 800 nm pump-probe experiments was observed in hole-doped cuprates across optimal doping, albeit with negative $\Delta R/ R$ at higher doping\cite{Gedik:AbruptTransitionOP}.  The sign change in hole-doped cuprates has been attributed to the plasma frequency shifting to higher frequency with doping\cite{Cooper:OpticsYBCOPlasmaFreq,Giannetti:UltrafastOpticsStronglyCorrelated} or to the photoexcited condensate shifting to lower frequencies with dopings\cite{Gedik:AbruptTransitionOP}.  In electron-doped cuprates, the plasma frequency likewise shifts to higher frequency with increased doping\cite{Onose:PG_NCCO}, which can produce a sign change opposite to the one observed; this favors a variant of the latter explanation.

The second observation about the isolated SC component is that it appears to have very little fluence dependence(Fig. \ref{Fig 3}(c)-(d)), and the small amount of residual fluence dependence may simply indicate that the condensate is not completely absent in the highest-field data which were used as the normal state reference.  The superfluid density is expected to decrease as $\sqrt{H}$ in \textit{d}-wave superconductors with line nodes\cite{Vekhter:QPsField_dSC,Luetkens:FieldDepSFDensity_FESC}, and this suggests an approximate remaining superfluid density equal to 12$\%$ (3 $\%$) of its zero field value for x$=$0.08 (x$=$0.11) LCCO measured at 11K (12K) and 6 T (6.5 T).  The weak or nonexistant fluence-dependence in Fig. \ref{Fig 3}(c)-(d) is in contrast to hole-doped cuprates which show pronounced fluence-dependence of the superconducting component, which was interpreted in terms of bimolecular recombination of quasiparticles, in several families\cite{Gedik:AbruptTransitionOP,Gedik:YBCOPRB,Mahmood:PrivateComm}.

Bimolecular recombination can dominate the initial recombination dynamics if either $\gamma_{pc}$ is small or if $\gamma_{esc}$ is large.  That is, if recombination is not balanced by pair breaking, either because the bosons formed from recombination have a small probability of breaking another Cooper pair or because these bosons escape from the excitation volume or decay into lower frequency excitations before they can break another Cooper pair. The relative magnitude of the different terms in the Rothwarf-Taylor equations does not necessarily imply a certain pairing mechanism.  For example, while underdoped hole-doped cuprates appear to have non-equilibrium dynamics in the superconducting state that are dominated by bimolecular recombination of quasiparticles, overdoped hole-doped cuprates (like electron-doped cuprates) show very little fluence dependence in the superconducting state\cite{Gedik:AbruptTransitionOP}, even though presumably the same pairing mechanism exists at all dopings.  Similarly, among BCS superconductors which have a phonon-mediated mechanism, fluence-dependence recombination rates have been observed in some systems\cite{Xi:MagneticFieldNbN_PP}.

\begin{figure}[!]
\includegraphics [type=eps,ext=.eps,read=.eps,clip, width=3 in]{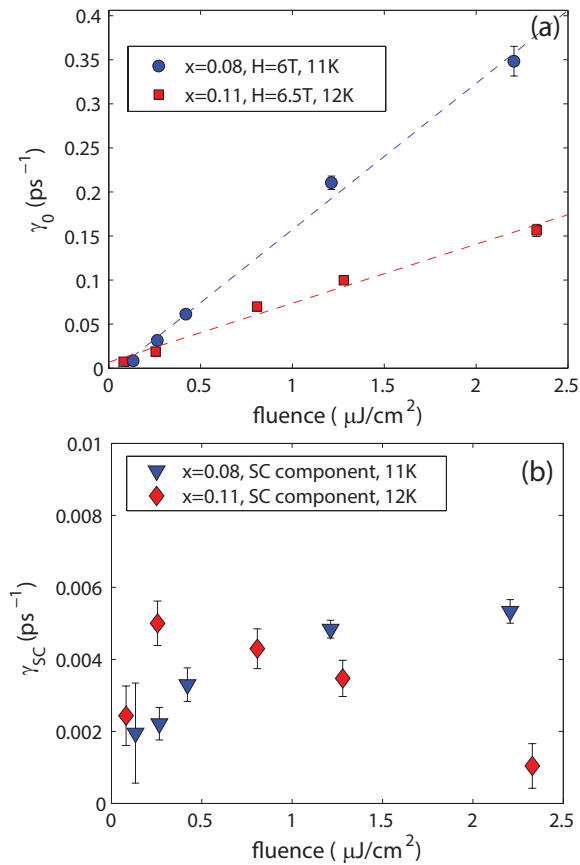}
\centering
\caption[Fig 4] {\label{Fig 4}  Initial decay rates as a function of pump fluence for AF (a) and SC (b) components for both dopings.  The former is fit to Egn. \ref{eqn fit bimolecular} and the latter is approximated by a linear slope (Fig. \ref{Fig 3}(d)). }
\end{figure}

The field-induced normal state (Fig. \ref{Fig 3}(e)-(f)) is marked by strong pump-fluence dependence of initial relaxation rates suggesting that the fluence dependence observed in the normal state (Fig. \ref{Fig 1}) persists to low temperature and needs to be accounted for when analyzing zero field data.  Fig. \ref{Fig 4} summarizes initial decay rates as a function of pump fluence for both the SC and the AF component at the measurement temperatures and fields in Fig. \ref{Fig 3}.  All decay rates represent a fit of data only after $\Delta R/R$ reaches its maximum deviation from zero.  Fig. \ref{Fig 4}(a) shows initial decay rates of the AF component, found by fitting to Eqn. \ref{eqn fit bimolecular}. Fig. \ref{Fig 4}(b) employs a linear approximation to extract the initial decay rate of the SC component ($\gamma_{SC}$), as indicated in Fig. \ref{Fig 3}(d).  The reason for the different fitting is because Eqn. \ref{eqn fit bimolecular} describes a bimolecular-recombination-dominated decay process which is likely not applicable to the SC component.  Fig. \ref{Fig 4} emphasizes the different fluence-dependent dynamics of decay across $\Delta_{AF}$ as compared to recombination of quasiparticles into Cooper pairs. The former depends strongly on fluence, varying systematically by more than an order of magnitude in the fluence regime examined.  The latter has little systematic fluence dependence, with the initial decay rate varying by only a factor of $\approx 3$, likely indicating a phonon bottleneck in the recombination.  This starkly different fluence-dependence may indicate different mechanisms of relaxation for the AF and SC excitations if $\gamma_{esc}$ is the rate-limiting parameter, but not necessarily if $\gamma_{pc}$ is responsible.

\begin{figure}[!]
\includegraphics [type=eps,ext=.eps,read=.eps,clip, width=3.3 in]{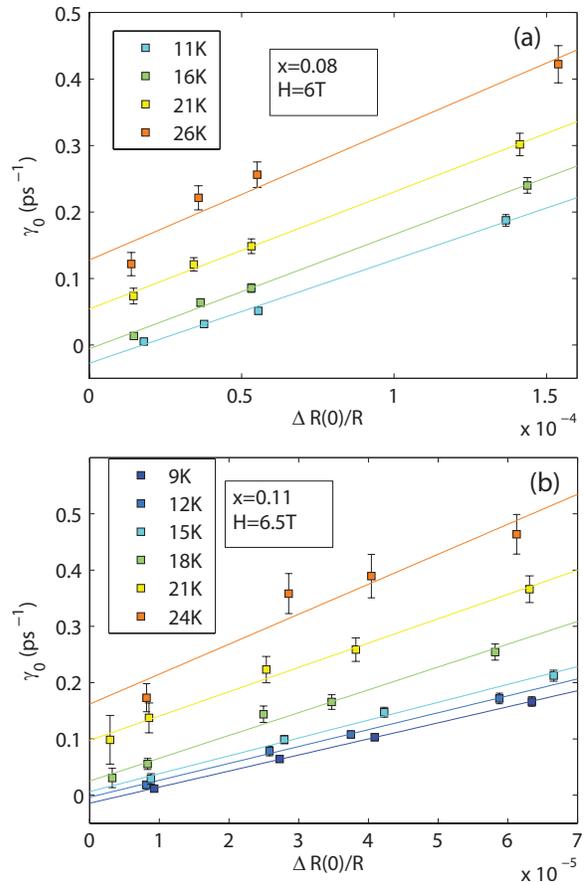}
\centering
\caption[Fig 5] {\label{Fig 5}  Temperature and fluence dependence of initial recombination rate in field-induced normal state.  Bimolecular recombination is implied by linear relationship between initial decay rates and magnitude of $\Delta R(0)/R$, which is proportional to fluence and number of excitations.  Fitting described in text. (a) x=0.08, taken at H=6T. (b) x=0.11, taken at H=6.5T }
\end{figure}
We explore the fluence and temperature dependence of initial decay rates in the field-induced normal state further in Fig. \ref{Fig 5}.  The key observation is that the fluence dependence at temperatures up to 26K is consistent with a bimolecular recombination process.  $|\Delta R(0)/R|$ is proportional to the fluence and the total number of excitations, and the initial recombination rate is derived by fitting $\Delta R/R(t)$ to Eqn. \ref{eqn fit bimolecular}.

At elevated temperatures, a linear relationship between $\gamma$ and $|\Delta R(0)/R|$ is maintained, though the data are uniformly shifted to higher values of $\gamma$.  This is consistent with increasing thermal populations ($n_{th}$) of excitations.  At elevated temperatures, thermally excited excitations can recombine with photoexcited excitations, giving a population evolution of photoexcited excitations of $\frac{dn_{ph}}{dt}=-\beta n_{ph}^2-\beta n_{ph}n_{th}$.  This expression yields a fluence dependence with the same slope, but with an offset given by $\gamma_{th}$, the recombination rate due to thermal excitations recombining with each other.  In the data, there is a small increase in slope as the temperature increases, particularly for $x=0.11$, and this is likely attributed to the smearing of the gap edge as temperature increases.  The result in Fig. \ref{Fig 5} indicates that the same bimolecular recombination process dictates the relaxation of the photoexcited state in the absence of superconductivity both at low temperature and at T$_c$.  Most likely, this bimolecular recombination happens across $\Delta_{AF}$.  As sketched in Fig. \ref{Fig 1a}, the gap is centered around $E_F$ only at the hot spots.  It should be noted that transient reflectivity can be sensitive to gaps not centered at $E_F$, such as the hybridization gap in heavy fermion materials\cite{Demsar:HF_PP_review}.  Another possibility is that the portions of the FS away from the hotspots simply contribute less to $\Delta R/R$.

\section{Discussion}

\subsection{Fluence dependence above T$_c$}
We begin by considering the appearance of fluence-dependent relaxation rates below a temperature consistent with T$_D$ from transport.  Because of this close correspondence in onset temperature, this characteristic fluence-dependence is attributed to AF correlations.  SC fluctuations are ruled out as an origin because those have been demonstrated to emerge at lower temperature\cite{Li:NernstPCCO} and because the isolated SC component lacks systematic fluence dependence.  It should be noted that the onset of AMR does not necessarily indicate long-range order or static order in LCCO, as $\mu SR$ experiments yield a much more truncated AF regime\cite{Saadaoui:LCCOmuSR}.  Previous theoretical work has been shown that decreasing AF correlation length has the effect of adding DOS into the gap produced by \textbf{q}$=$($\pi$,$\pi$) ordering\cite{Park:SpinCorrelationsCalculation,Nekrasov:ARPES_DMFT_PCCO}.  The DOS inside the gap has been related to the AF correlation length for the related compound Sm$_{2-x}$Ce$_x$CuO$_{4\pm\delta}$ (SCCO) \cite{Park:SpinCorrelationsCalculation}, and in-gap DOS starts to grow appreciably for correlation lengths smaller than 16 lattice constants.  In this context, the fluence-dependent relaxation rates are attributed to bimolecular recombination of electron-hole pairs across $\Delta_{AF}$ where the AF correlation length is sufficiently long to fully deplete the DOS inside the gap, as sketched in Fig. \ref{Fig 7}(c).  The temperature regime (T$>$T$_{D, AMR}$) without fluence dependence is attributed to AF with shorter correlation length such that there is a continuum of available states for hot electron thermalization.  Previously the onset of AF correlations in optical-pump probe data was identified via a second exponential term in the fitting\cite{Wang:2016UltrafastLCCO}, and we emphasize that the present identification via fluence dependence is apparent without fitting.

\subsection{Origin of bimolecular recombination}
When a particle and a hole undergo bimolecular recombination, a boson with the energy and momentum difference is created.  We attempt to identify this boson based on the energy scales involved.  Static optical conductivity has indicated that the magnitude of the gap due to AF correlations in electron doped cuprates\cite{Zimmers:SimulatingOptics,Armitage:ElectronDopedReview} is $2\Delta_{AF}\approx 18 k_b T^*$.  As shown in PCCO, $T^*$ as determined from optical conductivity tends to be higher than T$_D$ from AMR, which is within error bars of T$_N$ measured by optical pump-probe (Fig. \ref{Fig 1}).  Thus, our estimates of $2\Delta_{AF}$ of 85 meV and 124 meV for x$=$0.11 and x$=$0.08, respectively, based on transport and optical pump-probe values of T$_D$, represent lower bounds.

The first possibility is radiative recombination of electron-hole pairs with zero net momentum.  Second, we consider recombination mediated by optical phonons.  The highest frequency optical phonons in \textit{doped} electron-doped cuprates have been shown to have energy of $\approx 60$ meV\cite{dAstuto:OpticalPhononDisp_NCCO,Kang:DopingEvolutionPhononDOS_NCCO}, which is not sufficient to traverse $2\Delta_{AF}$.  It should be noted that in the regime of strong electron-phonon coupling, it is possible to have recombination be mediated by multiple phonons\cite{Werner:FieldInducedPolaronHolsteinHubbardUltrafastTheory}, though the dimensionless electron-phonon coupling constant, $\lambda$, in electron-doped cuprates is estimated to be $\approx 1$, in a moderately-coupled regime\cite{Park:E-Ph_coupling_variousElectronDopedCuprates_2008,Schmitt:AnalysisSpectralFuncNCCO_2008}.  Two other scenarios may allow for phonon-mediated recombination: a very broad gap edge, such that the effective lowest-energy excitation is smaller than the optical gap, and doping inhomogeneity, such that some regions of the sample have sufficiently small AF gap to be traversed by phonons. Both of these scenarios may be present in our samples.

The other possibility we consider are magnetic excitations--magnons or spin waves.  In undoped $PrLaCuO_4$ and lightly-doped NCCO, a magnon dispersion has been measured with a maximum energy of $\approx 300$ meV at \textbf{q}$=(\pi,\pi)$\cite{Ishii:HighEnergySpinChargeExcitations_electronDoped,Lee:AsymmetryCollectiveExcitationElectronDoped}.  With electron doping, the dispersion shifts to higher energy at all measured momentum transfer\cite{Ishii:HighEnergySpinChargeExcitations_electronDoped}, and near optimal doping, a dispersing collective mode with an energy of $\approx 300 meV$ at \textbf{q}$=(0,0)$ \cite{Lee:AsymmetryCollectiveExcitationElectronDoped}.  Bimolecular recombination of electrons and holes with net momentum $\mathbf{q}=0$ may be facilitated by the latter excitation.  Alternately, electrons and holes with momentum difference $\mathbf{q}=(\pi,\pi)$, such as those at opposite hot-spot momenta, may recombine facilitated by magnons.

\subsection{Connecting to physical quantities}
One point highlighted by the discrepancy between phase diagrams from $\mu SR$ and AMR experiments is that AF correlations exist on finite timescales.  Because these AF correlations are responsible for the gap in the DOS which gives rise to bimolecular recombination dynamics observed in the normal state, we can use these time-domain experiments to set limits on AF correlation times.  In particular, the recombination time in the limit of zero pump-fluence sets a lower bound on a characteristic timescale on which a given AF domain configuration is static.  This procedure gives a minimum AF correlation time at T$_c$ of $6.7\pm0.4$ ps and $1.9\pm0.3$ ps for x$=$0.08 and x$=$0.11, respectively.  At temperatures $T/T_c\approx0.6-0.7$, lower limits are placed at 35 ps(74 ps) from the lowest measured fluence of x$=$0.11(0.08).  The timescale on which AF correlations are static is useful for clarifying the relationship between AF and SC.  For example, if Cooper pairing happens on timescales and length scales shorter than AF correlations, a picture of microscopic coexistence, albeit only 2D, may still be reasonable in a regime where static 3D long-range order is absent.


\begin{figure}[!]
\includegraphics [type=eps,ext=.eps,read=.eps,clip, width=3.5 in]{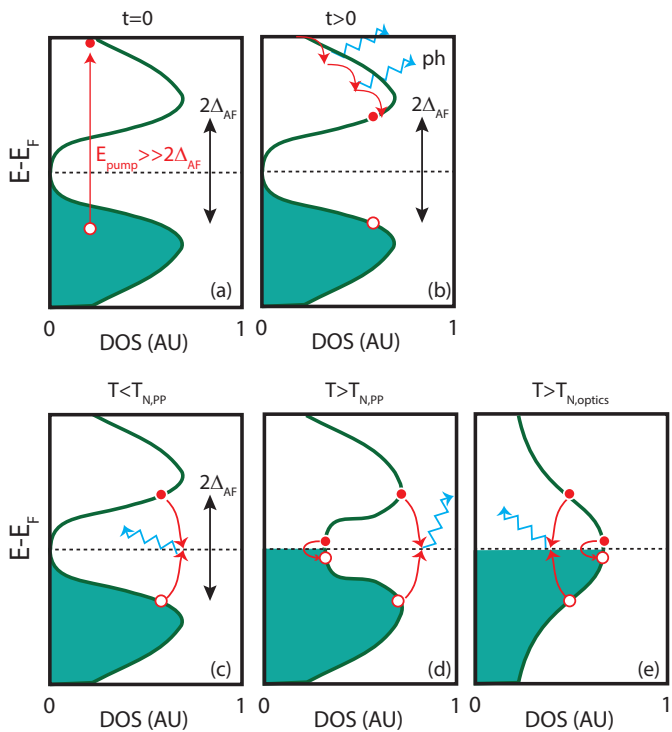}
\centering
\caption[Fig 7] {\label{Fig 7}  Summary. (a)-(b) Effect of the pump. (a) 1.5eV pump makes high energy excitations in excess of $\Delta_{AF}$. (b) High energy excitations decay to gap edge via emission of optical phonons or other high-energy bosons. (c)-(e) Relaxation in various temperature regimes. (c) When DOS inside gap is fully depleted, relaxation proceeds via bimolecular recombination across the gap, as evidenced by characteristic temperature and fluence dependence of initial decay rate, shown in Fig. \ref{Fig 5}. (d) When sufficient DOS are inside the gap, additional decay channels appear, and relaxation rates are no longer fluence dependent. (e) At sufficiently high temperature, corresponding to the pseudogap temperature in optics, the $\Delta_{AF}$ fills in completely. }
\end{figure}

The slope of the bimolecular recombination rate ($\propto \beta$) can be used to estimate the coupling between charged particles and bosonic excitations, analogous to how $\beta$ is related to the ratio of electron-phonon coupling to the DOS at $E_F$ in BCS superconductors\cite{Gedik:YBCOPRB}. In order to estimate a coupling parameter, we need an estimate of the number of excitations that are created for a given pump fluence.  As an upper bound, we assume that 100$\%$ of the pump fluence goes towards creating electron-hole excitations of energy $2\Delta_{AF}$.  Analysis of the rate of fluence dependence in the field-induced normal state at low temperature (Fig. \ref{Fig 4}) and above T$_c$ in a magnetic field (Fig. \ref{Fig 5} ) indicate that electron-boson coupling in x$=$0.08 is between 2 and 4 times stronger than in x$=$0.11, a much larger difference than the relative T$_c$'s of the two dopings.

\section{Conclusions}
We have performed optical pump-probe experiments on underdoped and optimally doped LCCO in the superconducting state, in the normal state above T$_c$, and in the field-induced normal state.  In the superconducting state, we observe distinct components in $\Delta R/R$ attributed to AF and SC.  The latter shows a sign change going from under- to optimal-doping, and also shows bottlenecked recombination dynamics.  This observation indicates either a high efficiency of pair-breaking by gap-frequency bosons and/or a slow diffusion and/or anharmonic decay rate of gap frequncy bosons.  The field-induced normal state, is marked by a strong fluence dependence of $\Delta R/R$.  The specific functional form of this fluence dependence suggests that the relaxation mechanism of  photoexcitations is dominated by bimolecular particle-hole recombination across a gap originating from AF correlations, and this recombination is likely mediated by magnetic excitations, as sketched in Fig. \ref{Fig 7}.  Fluence dependence persists to a temperature similar to T$_D$ observed by AMR, suggesting that the transport signature of AF correlations also corresponds to a fully depleted gap in the DOS (Fig. \ref{Fig 7}(c)).  Attributing the fluence-dependence of the normal state to bimolecular recombination allows us to put limits on both the timescale on which AF correlations are static and the strength of electron-magnon coupling.

\begin{acknowledgements}
We acknowledge helpful discussions with S. Lederer, T. Senthil, D. Torchinsky, and P. Werner.  Optical pump-probe work was supported by the Gordon and Betty Moore Foundation's EPiQS initiative through grant GBMF4540.  Materials growth and characterization was supported by AFOSR FA95501410332 and NSF DMR1410665.
\end{acknowledgements}

\bibliography{LCCO_refs}
\end{document}